\RequirePackage{lineno}
\documentclass[prl,twocolumn,showpacs,superscriptaddress,amsmath]{revtex4}
\usepackage{amssymb}
\usepackage{xspace}
\usepackage{graphicx}
\usepackage{dcolumn}
\usepackage{bm}
\usepackage{rotating}
\usepackage{color}
\usepackage{verbatim}
\usepackage{multirow}
\def \pp {\pi^+\pi^-}

\def \jp {J/\psi}

\def \ee {e^+e^-}
\def \uu {\mu^+\mu^-}

\def \zc {Z_c}
\def \qn {J^P}

\def \Flatte {Flatt\'{e}}

\def \ie {\emph{i.e.}}

\begin{document}

\title{\boldmath Determination of spin and parity of the $\zc(3900)$}

\author{
M.~Ablikim$^{1}$, M.~N.~Achasov$^{9,f}$, X.~C.~Ai$^{1}$, O.~Albayrak$^{5}$, M.~Albrecht$^{4}$, D.~J.~Ambrose$^{44}$, A.~Amoroso$^{49A,49C}$, F.~F.~An$^{1}$, Q.~An$^{46,a}$, J.~Z.~Bai$^{1}$
% ...
\\
\vspace{0.2cm}
(BESIII Collaboration)\\
\vspace{0.2cm} {\it
$^{1}$ Institute of High Energy Physics, Beijing 100049, People's Republic of China\\
% ...
$^{4}$ Bochum Ruhr-University, D-44780 Bochum, Germany\\
$^{5}$ Carnegie Mellon University, Pittsburgh, Pennsylvania 15213, USA\\
% ...
$^{9}$ G.I. Budker Institute of Nuclear Physics SB RAS (BINP), Novosibirsk 630090, Russia\\
% ...
$^{44}$ University of Rochester, Rochester, New York 14627, USA\\
% ...
$^{46}$ University of Science and Technology of China, Hefei 230026, People's Republic of China\\
% ...
$^{49}$ (A)University of Turin, I-10125, Turin, Italy; (B)University of Eastern Piedmont, I-15121, Alessandria, Italy; (C)INFN, I-10125, Turin, Italy\\
% ...
$^{a}$ Also at State Key Laboratory of Particle Detection and Electronics, Beijing 100049, Hefei 230026, People's Republic of China\\
% ...
$^{f}$ Also at the Novosibirsk State University, Novosibirsk, 630090, Russia\\
% ...
}
}

\begin{abstract}

The spin and parity of the $\zc(3900)^\pm$ state are determined to be
$J^P=1^+$ with a statistical significance larger than $7\sigma$ over other quantum numbers in a partial wave analysis
of the process $\ee\to \pp\jp$. We use a data sample of 1.92~fb$^{-1}$
accumulated at $\sqrt{s}=4.23$ and 4.26~GeV with the BESIII experiment.
When parameterizing the $\zc(3900)^\pm$ with a
\Flatte-like formula, we determine its pole mass
$M_\textrm{pole}=(3881.2\pm4.2_\textrm{stat}\pm52.7_\textrm{\rm syst})\textrm{~MeV}/c^2$
and pole width
$\Gamma_\textrm{pole}=(51.8\pm4.6_\textrm{stat}\pm36.0_\textrm{syst})\textrm{~MeV}$. We
also measure cross sections for the process $\ee\to\zc(3900)^+\pi^-+c.c.\to\jp\pp$
and determine an upper limit at the 90\% confidence level for the
process $\ee\to\zc(4020)^+\pi^-+c.c.\to\jp\pp$.

\end{abstract}


\maketitle

A charged charmoniumlike state, $\zc^\pm$ ($\zc$ denotes $\zc(3900)$
throughout this Letter except when its mass is explicitly mentioned), was observed by
the BESIII~\cite{bes3zc} and Belle~\cite{bell} collaborations in
the process $\ee\to \pp\jp$ and confirmed using CLEO-c's
data~\cite{cleoc}. As there are at least four quarks in the
structure, many theoretical interpretations of the
nature and the decay dynamics of the $\zc$ have been put forward \cite{theoryzc1,theoryzc2,theoryzc3,theoryzc4,theoryzc5,theoryzc6}.

A similar charged structure, the $\zc(3885)^\pm$, was observed in the
process $\ee\to (D\bar D^*)^\pm\pi^\mp$~\cite{xuxp}, with spin
parity ($J^P$) assignment of $1^+$ favored over the $1^-$ and $0^-$ hypotheses.
However, its mass and width are $2\sigma$ and $1\sigma$,
respectively, below those of the $\zc^\pm$ observed in $\ee\to\pp\jp$.
Are the $\zc(3885)^\pm$
and the $\zc^\pm$ the same state and do they have the same spin and
parity? This is one of the most important piece of information
desired in many theoretical
analyses~\cite{theoryzc3,braaten}.  Finally, the $\zc(4020)$ was
observed for the first time in the processes $\ee\to \pp h_c$~\cite{pipihc} and
$\ee\to (D^*\bar{D}^*)^\pm\pi^\mp$ \cite{ddpi}, but it has not been
searched for in the $\pp\jp$ final state yet.

In this Letter, we report on the determination of spin and parity of
the $\zc$ and a search for the $\zc(4020)^\pm$ in the process
$\ee\to\pp\jp$.  The results are based on a partial wave analysis
(PWA) of the $\ee\to \pp\jp$ events accumulated with the BESIII
detector~\cite{bes3}. The data sample includes 1092~$\rm pb^{-1}$
$e^+e^-$ collision data at a
center-of-mass (c.m.) energy $\sqrt{s}=4.23$~GeV, and
827~$\rm{pb}^{-1}$ data at $\sqrt{s}=4.26$~GeV~\cite{lumi}. The
precise c.m.\ energies are measured with the di-muon process
\cite{ecms}.

The $\ee\to \pp\jp$ candidate events are selected with the same
selection criteria as described in Ref.~\cite{bes3zc, eventslc} with $\jp$
reconstructed from lepton pairs ($\ell^+\ell^-=\uu,~\ee$). The
numbers of selected candidate events are 4154 at $\sqrt s=4.23$ GeV and
2447 at $\sqrt s=4.26$ GeV; the event samples are estimated to contain 365 and 272 background
events, respectively, at these two points, using the $\jp$ mass sidebands as has been done in
Ref.~\cite{bes3zc}.

Amplitudes of the PWA are constructed with the helicity-covariant
method~\cite{chung}; the process $\ee\to\pp\jp$ is assumed to
proceed via the $\zc$ resonance, \emph{i.e.}, $\ee\to \zc^\pm\pi^\mp$,
$\zc^\pm\to \jp\pi^\pm$, and via the non-$\zc$ decay $\ee\to R\jp$,
$R\to \pp$. All processes are added coherently to obtain the total
amplitude \cite{chenh}. For a particle decaying to the two-body final
state, \ie, $A(J,m)\to B(s,\lambda) C(\sigma,\nu)$, where spin and helicity are indicated in the parentheses, its helicity
amplitude $F_{\lambda,\nu}$ is related to the covariant amplitude
via~\cite{chung,vf}
\begin{equation}
\label{helamp}
 F_{\lambda, \nu}=\sum\limits_{lS} g_{lS}\sqrt{\frac{2l+1}{2J+1}}
 \langle l0S\delta|J\delta\rangle
 \langle s\lambda\sigma-\nu|S\delta\rangle r^{l}{B_{l}(r)\over~B_{l}(r_0)} ,
\end{equation}
where $\delta=\lambda-\nu$, and $g_{lS}$ is the coupling constant in the $l$-$S$ coupling
scheme, the angular brackets denote Clebsch-Gordan coefficients, $r$
is the magnitude of the momentum difference between the two
final state particles, $r_0$ corresponds to the momentum difference at
the nominal mass of the resonance, and
$B_l$ is a barrier factor~\cite{barrierform}. The
nonresonant process, $\ee\to \pp\jp$, is parameterized with an amplitude
based on the QCD multipole expansion~\cite{contactterm}.

The relative magnitudes and phases of the complex coupling constants
$g_{lS}$ are determined by an unbinned maximum likelihood fit to
data. The minimization is performed using the package {\sc minuit}~\cite{minuit}, and the
backgrounds are subtracted from the likelihood as in
Ref.~\cite{zhuc}.

In the nominal fit, we assume the $\zc$ to have $J^P=1^+$, and its lineshape is described with a \Flatte-like formula taking into
account the fact that the $\zc^\pm$ decays are dominated by the final states $(D\bar{D}^*)^\pm$~\cite{xuxp} and $\jp\pi^\pm$~\cite{bes3zc}, \ie,
\begin{equation}\label{zcflatte}
 BW(s,M,g'_1,g'_2)={1\over s-M^2+i[g'_1\rho_{1}(s)+g'_2\rho_{2}(s)]},
\end{equation}
where the subscripts in $g'_i~(i=1,2)$ represent the $\zc^\pm\to \pi^\pm\jp$ and $(D\bar{D}^*)^\pm$ decays, respectively; $\rho_{i}(s)=2k_{i}/\sqrt s$
is a kinematic factor with $k_{i}$ being the magnitude of the three-vector
momentum of the final state particle ($\jp$ or $D$) in the $\zc$ rest
frame; and $g'_1$ and $g'_2$ are the coupling strengths of $\zc^\pm\to
\pi^\pm\jp$ and $\zc^\pm\to (D\bar{D}^*)^\pm$, respectively, which will be
determined by the fit to data.

To describe the $\pp$ mass spectrum, four resonances, $\sigma$,
$f_0(980)$, $f_2(1270)$ and $f_0(1370)$, are introduced.
$f_0(980)$ is described with a \Flatte~formula~\cite{bes2gi}, and
the others are described with relativistic Breit-Wigner (BW)
functions. The width of the wide resonance $\sigma$ is
parameterized with
 $\Gamma_\sigma(s)=\sqrt{1-{4m^2_{\pi}\over s}}\Gamma$~\cite{berman,besiib},
and the masses and widths for the $f_2(1270)$ and $f_0(1370)$ are
taken from the Particle Data Group (PDG)~\cite{pdg}. The statistical
significance for each resonance is determined by examining the probability of
the change in log likelihood $(\log L)$ values between including and excluding this resonance in the fits, and the probability is calculated under the $\chi^2$ distribution hypothesis taking the change of the number of degrees of freedom
$\Delta({\rm ndf})$ into account. With this procedure, the statistical
significance of each of these
states and the nonresonant process is estimated to be larger than
5$\sigma$.  All of them are therefore included in the nominal fit, which includes the $\ee\to\sigma\jp$,
$f_0\jp$, $f_0(1370)\jp$, $f_2(1270)\jp$, $\zc^\pm\pi^\mp$ and nonresonant processes.

\begin{figure*}[htbp]
\vspace{0.3cm}
\includegraphics[width=0.95\textwidth]{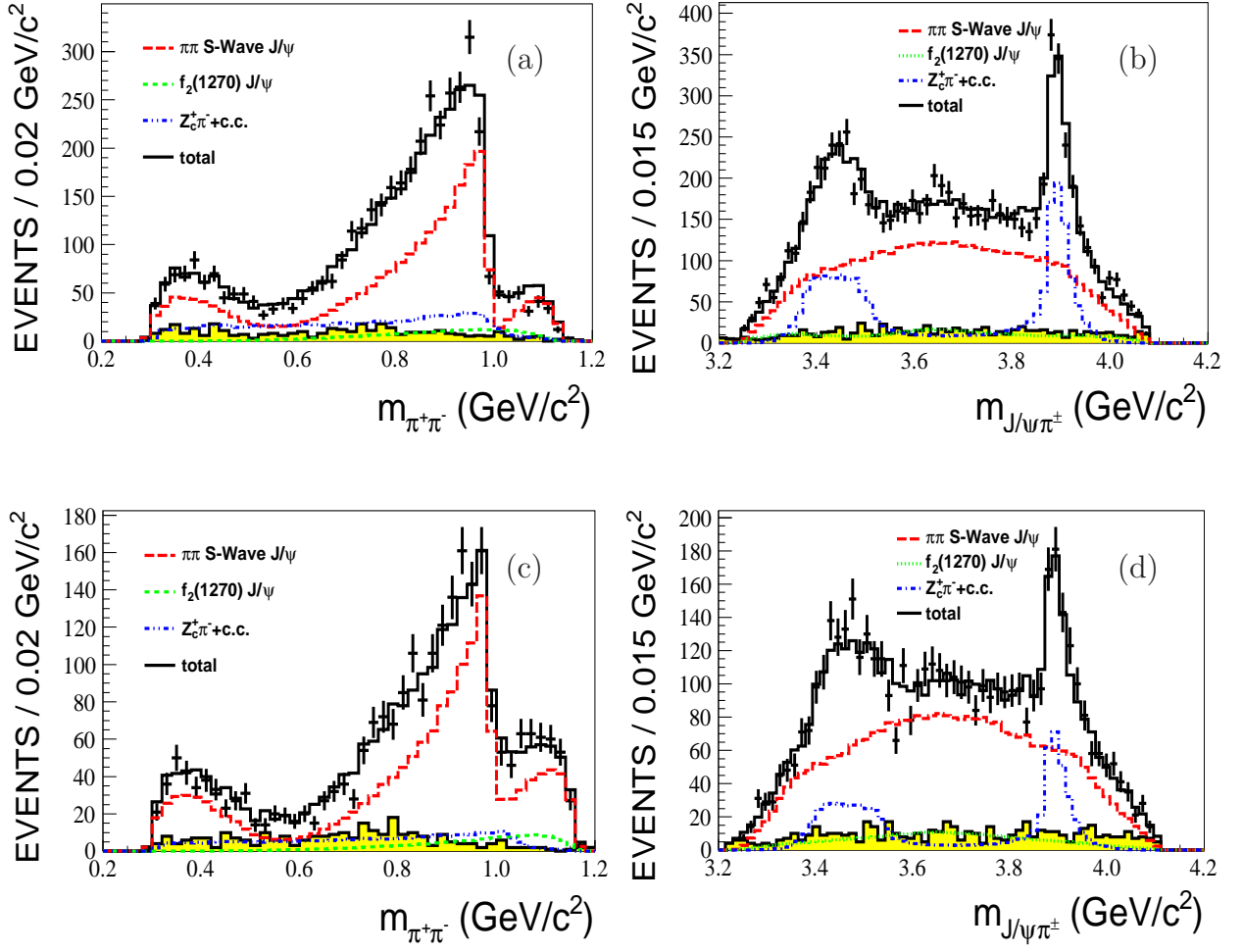}
\caption{\label{pwafitresult} (color online) Projections to $m_{\pp}$ (a,~c) and
$m_{\jp\pi^\pm}$ (b,~d) of the fit results with $J^P=1^+$ for the
$\zc$, at $\sqrt s=4.23$~GeV (a,~b) and $\sqrt s=4.26$~GeV
(c,~d). The points with error bars are data, and the black
histograms are the total fit results including backgrounds.  The shaded histogram denotes backgrounds.
The contributions from the $\pp~S$-wave $\jp$, $f_2(1270)\jp$, and $\zc^\pm\pi^\mp$,
are shown in the plots.  The $\pp~S$-wave resonances include the $\sigma$, $f_0(980)$ and $f_0(1370)$. Plots (b) and (d) are filled with two entries
($m_{\jp\pi^+}$ and $m_{\jp\pi^-}$) per event.}
\end{figure*}

A simultaneous fit is performed to the two data sets. The coupling constants are set as free parameters
and are allowed to be different at the two energy points except for the common ones describing $\zc$ decays. The oppositely charged $\zc$ states are
regarded as isospin partners; they share a common mass and
coupling parameters $g'_1$ and $g'_2$. Figure~\ref{pwafitresult}
shows projections of the fit results at $\sqrt s=4.23$ and 4.26~GeV. The mass of
$\zc^\pm$ is measured to be $M_{\zc}=(3901.5\pm
2.7_\textrm{stat})$~MeV/$c^2$ and the coupling parameters
$g'_1=(0.075\pm 0.006_\textrm{stat})$~GeV$^2$ and
$g'_2/g'_1=27.1\pm 2.0_\textrm{stat}$. This measurement is
consistent with the previous result $g'_2/g'_1=27.1\pm 13.1$
estimated based on the measured decay width ratio $\Gamma(\zc^{\pm}\to
(D\bar D^*)^{\pm})/\Gamma(\zc^{\pm}\to \jp\pi^{\pm}) = 6.2\pm
2.9$~\cite{xuxp}. If the $\zc^\pm$ is parameterized as a constant width BW
function, the simultaneous fit gives a mass of $(3897.6\pm
1.2_\textrm{stat})\textrm{~MeV}/c^2$ and a width of $(43.5\pm
1.5_\textrm{stat})\textrm{~MeV}$, but the value of $-\ln L$
increases by 22 with $\Delta(\text{ndf})=1$. The BW parametrization is
thus disfavored with a significance of 6.6$\sigma$.

\begin{figure*}[htbp]
\vspace{0.3cm}
\includegraphics[width=0.95\textwidth]{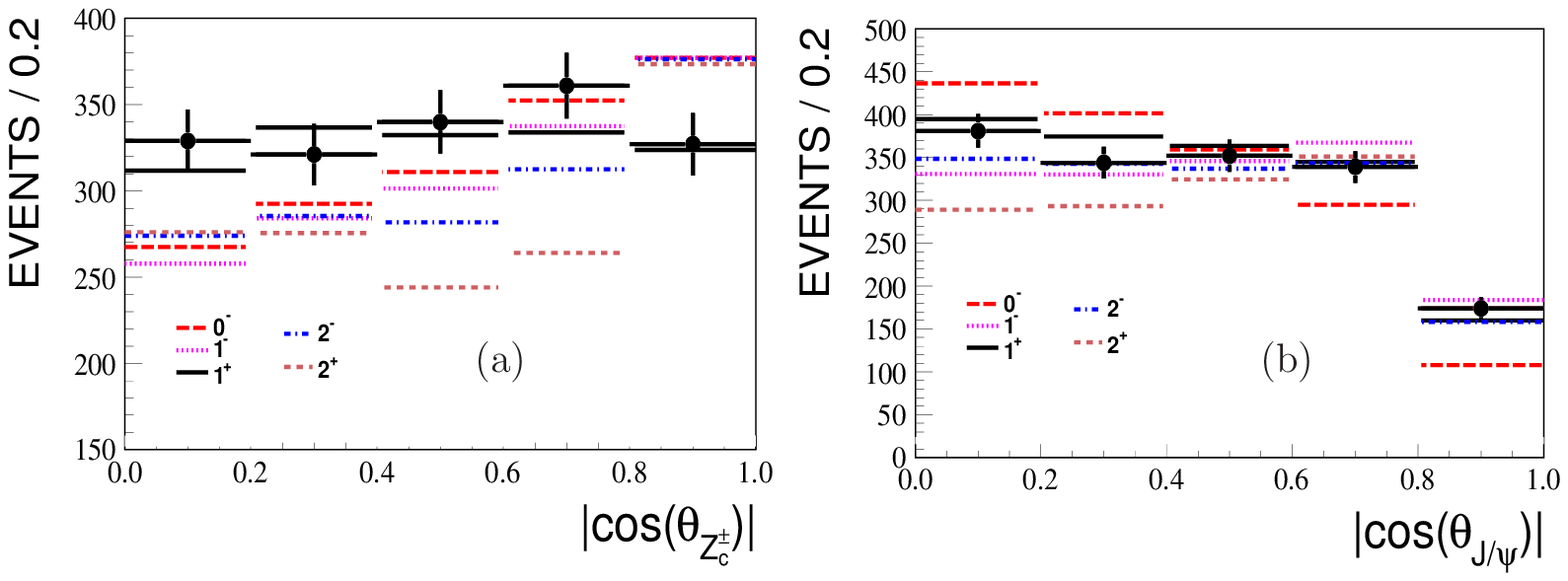}
\caption{\label{angdis} (color online) (a) Polar angle distribution of $\zc^\pm$ in the process $\ee\to\zc^+\pi^- + c.c.$, (b) helicity angle distribution of $\jp$ in the $\zc^\pm\to\pi^\pm\jp$. The dots with error bars show the combined data with requirement $m_{\jp\pi^\pm}\in(3.86,3.92)$ GeV/$c^2$, and compared to the total fit results with different $J^P$ hypotheses.}
\end{figure*}

Figure \ref{angdis} shows the polar angle ($\theta_{\zc^\pm}$)
distribution of $\zc^\pm$ in the process $\ee\to\zc^+\pi^- + c.c.$
and the helicity angle $(\theta_{\jp})$ distribution in the decay
$\zc^\pm\to\pi^\pm\jp$ for the combined data within the $\zc$ mass
region $m_{\jp\pi^\pm}\in(3.86,3.92)$~GeV/$c^2$, where $\theta_{\jp}$ is the
angle between the momentum of $\jp$ in the $\zc$ rest frame and the
$\zc$ momentum in the $\ee$ rest frame. The fit results, using
different assumptions for the $\zc$ spin and parity, are drawn with a
global normalization factor. The distribution indicates that data
favors a spin and parity assignment of $1^+$ for the $\zc^\pm$.
The significance of the $\zc^\pm(1^+)$ hypothesis is further examined
using the hypothesis test~\cite{cl}, in which
the alternative hypothesis is our nominal fit with an additional
$\zc^\pm(J^P\neq1^+)$ state. Possible $J^P$ assignments, other than
$1^+$, are $0^-$, $1^-$, $2^-$, and $2^+$. The changes $-2\Delta\ln L$
when the $\zc(1^+)\pi^\mp$ amplitude is removed from the alternative
hypothesis are listed in Table~\ref{sign}. Using the associated change
in the ndf when the $\zc^\pm(1^+)$ is excluded, we determine the
significance of the $1^+$ hypothesis over the alternative $J^P$ possibilities to be larger than 7$\sigma$.

\begin{table}
\begin{center}
\caption{Significance of the spin parity $1^+$ over other quantum numbers for
$\zc^\pm$. The significance is obtained for given change in ndf,
$\Delta({\rm ndf}$). In each case, $\Delta({\rm ndf)}=2\times4+5$,
where $2\times 4$ ndf account for the coupling strength for $\ee\to
\zc^\pm\pi^\mp$ at the two data sets, and the additional five ndf are
the contribution of the common degrees of freedom for the $\zc$
resonant parameters and the coupling strength for $\zc^\pm\to \jp\pi^\pm$. \label{sign}}
\begin{tabular}{cccc}
\hline\hline Hypothesis   & $\Delta(-2\ln L)$ & $\Delta({\rm ndf)}$ & Significance \\\hline
$1^+$ over $0^-$ & 94.0   &13& 7.6$\sigma$\\
$1^+$ over $1^-$ & 158.3  &13& $ 10.8\sigma$\\
$1^+$ over $2^-$ & 151.9  &13& $10.5\sigma$\\
$1^+$ over $2^+$ & 96.0   &13& $ 7.7\sigma$\\
\hline\hline
\end{tabular}
\end{center}
\end{table}

The fit results shown in Fig.~\ref{pwafitresult} indicate that process is dominated by the
$\pi\pi~S-$wave resonances, {\it i.e.} the $\sigma$, $f_0(980)$ and $f_0(1370)$. The
fraction of all $\pp$ $S$-wave components including the
interference between them is measured to be $(61.7\pm 2.1_\textrm{stat})$\% of
the total $\pp\jp$ events at $\sqrt{s}=4.23$~GeV and $(71.4\pm
4.1_\textrm{stat})\%$ at $\sqrt{s}=4.26$~GeV. The signal yields $N_{\zc^\pm}$ of $\zc^\pm$ are calculated by scaling its partial signal ratio with the total number of signal events. They are measured to be $N_{\zc^\pm}=952.3\pm 39.3_\textrm{stat}$ at $\sqrt s=4.23$~GeV and
$343.3\pm 23.3_\textrm{stat}$ at $\sqrt s=4.26$~GeV. Here, the errors are
statistical only, and they are estimated using the covariance matrix
from the fits.

To measure amplitudes associated with the polarization of $\zc^\pm$ in $\ee\to \zc^\pm\pi^{\mp}$ and
that of $\jp$ in $\zc^{\pm}\to \jp\pi^\pm$ decays in the nominal fit, the ratios of helicity amplitudes with different polarizations as defined in Eq.~\eqref{helamp} are calculated to be $|F^{\zc}_{1,0}|^2/|F^{\zc}_{0,0}|^2=0.22\pm 0.05_\textrm{stat} $ at 4.23 GeV, and $0.21\pm0.11_\textrm{stat}$ at 4.26 GeV  for
$\ee\to\zc^\pm\pi^{\mp}$, and $|F^{\psi}_{1,0}|^2/|F^{\psi}_{0,0}|^2=0.45\pm
0.15_\textrm{stat}$ for $\zc^\pm\to\jp\pi^{\pm}$, at both energy
points. Here $F^{\zc/\psi}_{1,0}$ and $F^{\zc/\psi}_{0,0}$ correspond to transverse and
longitudinal polarization amplitudes in the decay, respectively. The results show that the $\zc$ polarization is dominated by the longitudinal component.

The Born cross section for $\zc$ production is measured with the
relation  $\sigma=N_{\zc^\pm}/(\mathcal{L}(1+\delta)\epsilon \mathcal{B})$,
where $N_{\zc^\pm}$ is the signal yield for the process $\ee\to
\zc^+\pi^-+c.c.\to \pp\jp$, $\mathcal{L}$ is the integrated
luminosity, and $\epsilon$ is the detection efficiency obtained from a
MC simulation which is generated using the amplitude parameters
determined in the PWA. The radiative correction factor $(1+\delta)$
is determined to be 0.818~\cite{bes3zc}. The Born
cross section is measured to be
 $(22.0\pm 1.0_\textrm{stat})$~pb at $\sqrt s=4.23$~GeV
 and
 $(11.0\pm 1.2_\textrm{stat})$~pb at $\sqrt s=4.26$~GeV.

Using these two data sets, we also search for the process
$\ee\to \zc(4020)^+\pi^-+c.c.\to \pp\jp$, with the $\zc(4020)^\pm$
assumed to be a $1^+$ state. In the PWA, its mass is taken from Ref.~\cite{pipihc},
and its width is taken as the
observed value, which includes the detector resolution. The
statistical significance for $\zc(4020)^\pm \to\jp\pi^\pm$ is found
to be 3$\sigma$ in the combined data. The Born cross sections are
measured to be $(0.2\pm0.1_\textrm{stat})$~pb at $4.23$ GeV and
$(0.8\pm0.4_\textrm{stat})$~pb at $s=4.26$~GeV, and the corresponding upper limits at
the 90\% confidence level are estimated to be $0.9$~pb and $1.4$~pb, respectively.

Systematic errors associated with the event selection, including
the luminosity measurement, tracking efficiency of charged tracks,
kinematic fit, initial state radiation (ISR) correction factor and
the branching fraction of $Br(\jp\to \ell^+\ell^-)$, have been estimated
to be 4.8\% for the cross section measurement and 1.8 MeV for the
$\zc$ mass in the previous analysis~\cite{bes3zc}.

Uncertainties associated with the amplitude analysis come from the
$\sigma$ and $\zc$ parametrizations, the background estimation,
the parameters in the $f_0(980)$ \Flatte~formula, the barrier
radius in the barrier factor, the mass resolution and the
component of non-resonant amplitude.

The systematic uncertainty due to the $\sigma$ lineshape is
estimated by comparing the nominal fit with two other
parameterizations,
the PKU~ansatz~\cite{pku} and the Zou-Bugg approach~\cite{zb}. The
differences in the $\zc$ signal yields and mass measurement are
taken as the errors, which are 2.5\% (31.0\%) for the signal yields
at 4.23 (4.26)~GeV and 19.5 MeV for the $\zc$ mass.

The uncertainty due to the $f_0(980)$ lineshape is estimated by
varying the couplings by 1$\sigma$ as determined in the decays
$\jp\to \phi\pp$ and $\phi K^+ K^-$~\cite{bes2gi}. Uncertainties
associated with the $f_0(1370)$ are estimated by varying the mass
and width by one standard deviation around the world average
values~\cite{pdg}.

The uncertainty due to the $\zc$ parametrization is estimated by
using a constant-width relativistic BW function.
The simultaneous fit gives the $\zc$ mass of $(3897.6\pm
1.2_\textrm{stat})$~MeV/$c^2$ and the width of $(43.5\pm
1.5_\textrm{stat})$~MeV. The difference in the $\zc$ signal
yields is 15.5\% (7.9\%) for the data taken at 4.23 (4.26)~GeV.

The uncertainty due to the background level is estimated by changing
the number of background events by $1\sigma$ around the nominal value,
that is, $\pm 25$ around 637 events.

The barrier radius is usually taken in the range $r_0\in
(0.25,~0.76)$~fm, with 0.6~fm being used in the nominal fit.
Uncertainties at both ends are checked. For a conservative
estimation, the radius $r_0=0.76$~fm,  which results in the larger difference, is used to estimate the
uncertainty.

The uncertainty due to the mass resolution in the $\jp\pi$ invariant mass
is estimated with an unfolded $\zc$ width. A truth width is
unfolded from the observed $\zc$ width using a relation determined by the MC simulation, and its difference
from the unfolded width, $\delta\Gamma/\Gamma=\delta g_1'/g_1'$, is taken as the systematic uncertainty for the coupling
constant $g_1'$. The uncertainties in the signal yields and the $\zc$ mass are determined with the truth coupling
constant.

The nonresonant process is described with a formula derived from the QCD multipole
expansion~\cite{contactterm}. It includes the $S$- and $D$-wave
components. The uncertainty associated with this amplitude is
estimated by removing the insignificant $D$-wave component and using
the $S$-wave component only.

\begin{table*}[htbp]
\caption{Summary of systematic uncertainties on the $\zc~(\qn=1^+)$
  mass, parameters $g_1'$ and $g_2'$, and the signal yields at 4.23
  GeV ($N_{\zc}^\textrm{I}$) and 4.26 GeV ($N_{\zc}^\textrm{II}$). The
  uncertainties shown for the $\zc$ mass, parameter $g_1'$ and the
  ratio $g_2'/g_1'$ are absolute values, while the uncertainties for $N_{\zc}^\textrm{I}$ and $N_{\zc}^\textrm{II}$ are relative ones.     \label{syserr}}
\begin{center}
\begin{tabular}{lccccc}
\hline\hline
Sources& $\zc$ Mass (MeV/$c^2$)& $g_1'\times10^{3}$ (GeV$^2$)& $g_2'/g_1'$&$N_{\zc}^\textrm{I}$ (\%)& $N_{\zc}^\textrm{II}$ (\%)  \\\hline
Event selection            &1.8   &...  &...&4.8&4.8 \\
$\sigma$ lineshape         &19.5  &12.0 &0.3&2.5&31.0\\
$\zc$ parametrization      &3.9   &...  &...&15.5&7.9\\
Backgrounds                &13.9  &8.0  &0.1&1.9&9.3\\
$f_0(980),g_1,g_2/g_1$     &17.5  &14.0 &0.6&2.4&24.6\\
$f_0(1370)$                &16.7  &11.0 &0.4&11.5&14.0\\
Barrier radius             &7.9   &2.0  &1.7&0.5&12.9\\
$\zc$ mass resolution      &1.0   &2.0  &...&0.4&0.5\\
Nonresonance              &14.3  &9.0  &0.0&0.1&18.0\\\hline
Total                      &38.0  &24.8 &1.9&20.3&49.2\\
\hline \hline
\end{tabular}
\end{center}
\end{table*}


Table \ref{syserr} summarizes the systematic uncertainties. Assuming
all of these sources are independent, the total systematic
uncertainties are 38.0 MeV for the measurement of the $\zc$ mass, and 20.3\%
(49.2\%) for the measurement of $\zc$ cross sections at
$\sqrt{s}=4.23$~(4.26)~GeV.

In summary, with 1.92~fb$^{-1}$ data taken at $\sqrt s=4.23$ and
4.26~GeV, the $\zc^\pm$ state is studied with an amplitude
fit to the $\ee\to \pp\jp$ samples, and its spin and parity have been determined to
be $1^+$ with a statistical significance larger than 7$\sigma$ over other quantum numbers. The mass is measured to be $M_{\zc}=(3901.5\pm
2.7_\textrm{stat}\pm 38.0_\textrm{syst})$~MeV$/c^2$ in the parametrization of a \Flatte-like
formula with parameters $g'_1=0.075\pm
0.006_\textrm{stat}\pm0.025_\textrm{syst}$~GeV$^2$, and $g'_2/g'_1=27.1\pm
2.0_\textrm{stat}\pm1.9_\textrm{syst}$, which corresponds to the $\zc$ pole mass $M_\text{pole}=(3881.2\pm4.2_\textrm{stat}\pm52.7_\text{syst})\text{~MeV}/c^2$ and pole width $\Gamma_\textrm{pole}=(51.8\pm4.6_\textrm{stat}\pm36.0_\textrm{syst})\textrm{~MeV}$, where $M_\textrm{pole}-i\Gamma_\textrm{pole}/2$ is the solution for which the denominator of \Flatte-like
formula is zero. The pole mass is consistent with the previous measurement \cite{xuxp}. The Born cross sections for the process $\ee\to\pi^+\zc^-+c.c.$
are measured to be
 $(21.8\pm 1.0_\textrm{stat}\pm 4.4_\textrm{syst})$~pb at $\sqrt s=4.23$~GeV
 and
 $(11.0\pm 1.2_\textrm{stat}\pm 5.4_\textrm{syst})$~pb at $\sqrt s=4.26$~GeV.
The contributions from $\zc(4020)^\pm$ are also searched for, but no
significant signals are observed, and an upper limit for the $\ee\to\pi^+\zc(4020)^-+c.c.$ process is determined to be 0.9 (1.4) pb at $\sqrt s=4.23~(4.26)$ GeV.

The BESIII collaboration thanks the staff of BEPCII and the
computing center for their strong support. This work is supported
in part by the Ministry of Science and Technology of China under
Contract No. 2009CB825200; Joint Funds of the National Natural
Science Foundation of China under Contracts Nos. U1332201;
National Natural Science Foundation of China (NSFC) under
Contracts Nos. 11175188, 11375205, 11235011, 11375221, 11565006, 10825524; German Research Foundation DFG
under Contract No. Collaborative Research Center CRC-1044, 627240;
Istituto Nazionale di Fisica Nucleare, Italy;  Ministry of
Development of Turkey under Contract No. DPT2006K-120470; U.S.
Department of Energy under Contracts Nos. DE-SC-0012069,
DE-SC-0010504, DE-SC-0010118, DE-FG02-05ER41374; U.S. National
Science Foundation; University of Groningen (RuG) under Contracts No. 530-4CDP03, and the
Helmholtzzentrum fuer Schwerionenforschung GmbH (GSI), Darmstadt;
WCU Program of National Research Foundation of Korea under
Contract No. R32-2008-000-10155-0.

\end{document}